\documentclass[epj]{webofc}
\usepackage[utf8]{inputenc}
\usepackage[varg]{txfonts}   
\usepackage{booktabs}
\usepackage{xcolor}
\definecolor{darkred}{rgb}{0.4,0.0,0.0}
\definecolor{darkgreen}{rgb}{0.0,0.4,0.0}
\definecolor{darkblue}{rgb}{0.0,0.0,0.4}
\usepackage[bookmarks,linktocpage,colorlinks,
    linkcolor = darkred,
    urlcolor  = darkblue,
    citecolor = darkgreen]{hyperref}
\usepackage{subfigure}
\wocname{EPJ Web of Conferences}
\woctitle{Lattice2017}
\newcommand{\tr}{\text{Tr\,}}
\begin{document}
%
\selectlanguage{english}
\title{%
Spectrum of QCD at Finite Isospin Density
}
\author{%
\firstname{Philipp} \lastname{Scior}\inst{1}\fnsep\thanks{Speaker, \email{scior@uni-muenster.de}},
\firstname{Lorenz} \lastname{von Smekal}\inst{2} \and
\firstname{Dominik} \lastname{Smith}\inst{2}
}
\institute{%
Institut für Theoretische Physik, Westfälische Wilhelms-Universität, Münster, Germany
\and
Institut für Theoretische Physik, Justus-Liebig-Universität, Gießen, Germany
}
\abstract{%
  We study the phase diagram of QCD at finite isospin density using two flavors of staggered quarks. We investigate the low temperature region of the phase diagram where we find a pion condensation phase at high chemical potential. We started a basic analysis of the spectrum at finite isospin density. In particular, we measured pion, rho and nucleon masses inside and outside of the pion condensation phase. In agreement with previous studies in two-color QCD at finite baryon density we find that the Polyakov loop does not depend on the density in the staggered formulation.
}
\maketitle
\section{Introduction}\label{intro}
Knowledge about the phase diagram of QCD is of great importance for studying various phenomena in nature. Unfortunately, lattice QCD is unable to provide us with results for the bigger part of the phase diagram as lattice QCD suffers from the fermion sign problem at finite baryon chemical potential \cite{Dzhunushaliev2016}. Yet, there remains another interesting region of the phase diagram that can be explored with standard lattice techniques, as there is no sign problem in this region: the part of the phase diagram with finite isospin density. By studying the finite isospin density region of the QCD phase diagram we can learn about general features of relativistic field theories at finite density that also occur in QCD with finite baryon density, e.g. the Silverblaze property \cite{Cohen:2003kd}. We may also use lattice studies in this region to learn about lattice cutoff effects at finite density, like saturation effects, and how to disentangle those effects from the relevant continuum physics.\\
On the other hand, the finite isospin density region of the QCD phase diagram is interesting by itself, as there are many systems in nature where we find an imbalance of neutrons and protons, leading to a finite isospin density. Such an imbalance occurs in nuclei, where there are more neutrons than protons, neutron stars, heavy ion collisions and possibly the early universe \cite{Schwarz2009}. Another intriguing feature that might not be of relevance for the aforementioned systems but has close analogies in ultracold fermionic quantum gases \cite{Boettcher2015} is the condensation of charged pions in the ground state of finite isospin density QCD. It is known from earlier studies \cite{PhysRevLett.86.592} that at low temperatures and high isospin chemical potential we find a second order phase transition to a phase with a Bose-Einstein condensate of charged pions.\\
In this contribution we concentrate particularly on the spectrum of QCD and how the masses of bound states change with varying isospin chemical potential. We investigate the masses of light mesons and the nucleons outside and inside the pion condensation phase at low temperature.

\section{Setup}\label{sec-1}
We use the tree-level Symanzik improved gauge action and staggered quarks. Our fermion matrix is
\begin{equation}
	M = \begin{pmatrix}
	D(\mu) & \lambda \eta_5 \\
	- \lambda \eta_5 & D(-\mu)
	\end{pmatrix}\, ,
	\label{eq-1}
\end{equation}
where $D$ is the staggered Dirac operator coupled to the up- and respectively the down-quark chemical potential, $\lambda$ is the the real coefficient of a pion source term that we introduce to be able to compute the charged pion condensate. Further, the pion source term also acts as a infrared cutoff - similar to a twisted mass term - that enables us to invert the Dirac operator in the pion condensation phase. Though its introduction is necessary for computational reasons we have to extrapolate $\lambda \to 0$ to extract physical results \cite{Brandt2016}. Finally we employ rooting to have $N_f=2$ degenerate quark flavors.

\section{Phase Diagram}\label{sec:phasediagram}
\begin{figure}[thb]
\centering
\includegraphics[width=0.5 \textwidth]{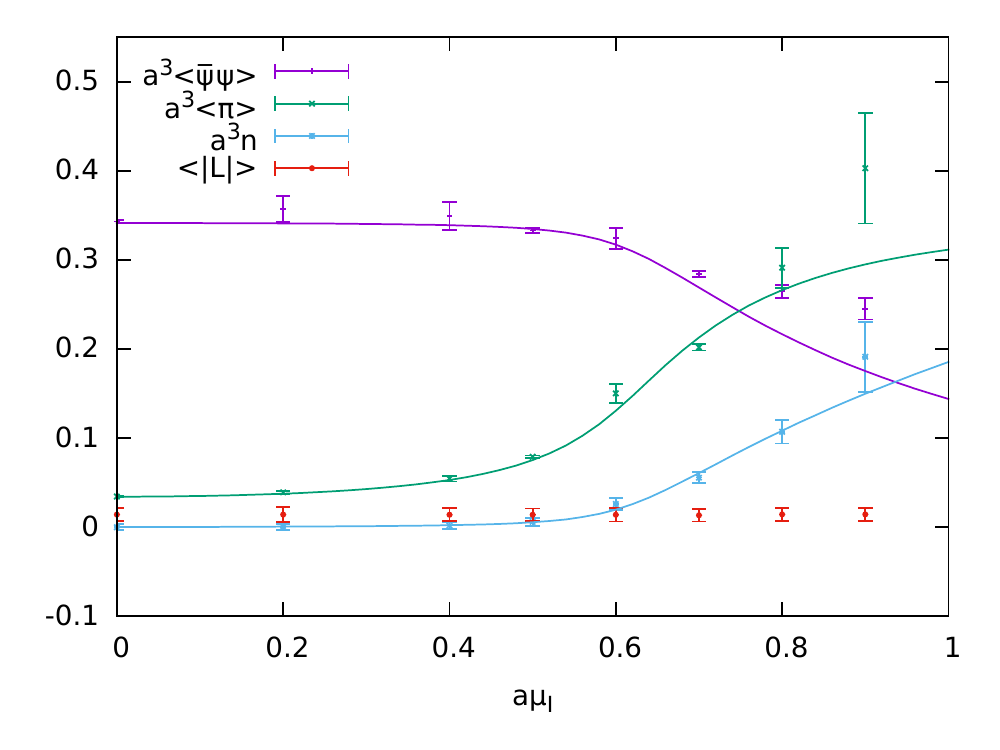}
\caption{Bulk thermodynamic observables for the parameter set $N_s=16$, $N_t=24$, $\beta=4.0$, $m=0.05$, $\lambda=0.005$ and several values of $\mu_I$. The solid lines are predictions by chiral perturbation theory.}
\label{fig-1}
\end{figure}\noindent A good starting point to evaluate, if the algorithms work is the calculation of bulk thermodynamic observables in the $(T,\mu_I)$-plane of the phase diagram. Previous studies that explored the phase diagram of QCD with finite isospin density can be found in \cite{PhysRevLett.86.592, Detmold2012, Kogut2002-1,Kamikado:2012bt, Carignano2017, Brandt2016}. To test our setup we have calculated the Polyakov loop $\langle |L|\rangle$, chiral condensate $\langle \bar \psi \psi \rangle$, charged pion condensate $\langle \pi \rangle$ and the net isospin density $n$ for a fixed temperature with the parameter set $N_s=16$, $N_t=24$, $\beta=4.0$, $m=0.05$, $\lambda=0.005$ for several values of isospin chemical potential. Figure~\ref{fig-1} shows the results for the previously mentioned observables together with predictions from chiral perturbation theory \cite{Kogut2000477}. We find that all thermodynamic observables stay constant at their vacuum values until at some value of $\mu_I$ the pion condensate and the net isospin density start to rise and the chiral condensate starts to drop. This indicates that we entered the phase where the ground state is a Bose-Einstein condensate of charged pions. The agreement of our lattice data with the lines from chiral perturbation theory is good up to high values of $\mu_I$ deep inside the pion condensation phase. Therefore, we conclude that our algorithms work fine. Another point worth mentioning is that the Polyakov loop stays at its vacuum value independent of $\mu_I$. This is in contrast to earlier works on QCD-like theories at finite density with Wilson fermions \cite{Cotter2012}. However it seems to be the generic case when using staggered fermions \cite{Holicki2016, Braguta2016}. The difference between both fermion discretizations at finite density should be explored by further studies in the future.

\section{Spectroscopy}
\begin{table}
\small
\centering

\begin{tabular}{l c c r }
			\hline
  			& Operator & $J^{PC}$ & Particle \\ \hline
  			I & $\eta_4 \bar \chi \chi$ & $0^{++}$ & - \\
  		   	& & $0^{-+}$ & $\pi$ \\
  			II&$\eta_i \eta_5 \zeta_i \bar \chi \chi$ & $1^{++}$ & $a_1$ \\
  		 	& & $1^{--}$ & $\rho$ \\
  			III&$\eta_i \eta_5 \eta_4 \zeta_i \zeta_4 \bar \chi \chi$ & $1^{+-}$ & $b_1$ \\
  		 	& & $1^{--}$ & $\rho$ \\
  			IV &$\epsilon_{abc}\chi_a \chi_b \chi_c$ & $\frac12^{-}$ & N(1535) \\
  		  		& & $\frac12^{+}$ & N \\ \hline 
		\end{tabular}
\caption{Operators, quantum number and particle identification for local irreducible particle representations. Space-time arguments are omitted.}
\label{tab-1}
\end{table}
As we are using staggered fermions, we have to be careful in choosing our lattice operators corresponding to the physical states in the continuum limit \cite{Altmeyer1993}. For our analysis of the spectrum at finite isospin density we use the four operators state in table~\ref{tab-1}. In particular we are interested in measuring the masses of the pions, the rho mesons and the nucleons. The pion masses can be easily extracted by measuring time-like correlators of operator I. Extracting the rho and nucleon masses, we have to be a little more careful, as operators II-IV will always contain a mixture of continuum states with different parities. Thus, we choose two lattice operators containing the rho meson to get some sense of our systematic uncertainties.

\subsection{Spectroscopy with finite pion sources} 
Introducing a finite pion source term $\frac{\lambda }{2}(\chi^T \tau_2 \chi + \bar \chi \tau_2 \bar \chi^T)$ in the Lagrangian, where $\tau_2$ is the second Pauli matrix in isospin space, changes the particle correlators significantly. Not only does it act as an infrared regulator for the eigenvalues of the Dirac operator, the introduction of the source term also leads to the appearance of non-diagonal parts of the fermion matrix, see equation~\eqref{eq-1}. This is of course also true for the fermion propagator
\begin{equation}
	G = M^{-1} = \begin{pmatrix}
				(D^\dagger_\mu D_\mu +\lambda^2)^{-1}D^\dagger_\mu & - \eta_5 (D^\dagger_\mu D_\mu +\lambda^2)^{-1} \lambda \\
				\eta_5 (D^\dagger_{-\mu} D_{-\mu} +\lambda^2)^{-1} \lambda & (D^\dagger_{-\mu} D_{-\mu} +\lambda^2)^{-1}D^\dagger_{-\mu}
				\end{pmatrix} \, .
\end{equation}
Thus, an up quark can propagate into a down quark and vice versa and we find disconnected pieces contributing to the pion and rho correlators, even though they are iso-triplet states. In the correlator for the neutral mesons we further find additional terms also in the connected contributions. The pion correlators are given by
\begin{align}
	\langle \pi^+(t) \bar \pi^+(0) \rangle =& -\eta_4(t) \tr_c [G_{11}(t,0,\mu) G_{11}(0,t,-\mu)] + \mathcal{O}(\lambda^2)\, , \notag\\
	\langle \pi^-(t) \bar \pi^-(0) \rangle =& -\eta_4(t) \eta_5(t) \tr_c [G_{22}(t,0,\mu) G_{22}^\dagger(t,0,\mu)]  + \mathcal{O}(\lambda^2)\, ,\\
    \langle \pi^0(t) \bar \pi^0(0) \rangle =& -\eta_4(t) \eta_5(t) \tr_c [G_{22}^\dagger(t,0,\mu) G_{11}(t,0,\mu) - G_{21}^\dagger(t,0,\mu) G_{21}(t,0,\mu) \notag\\
    &- G_{12}^\dagger(t,0,\mu) G_{12}(t,0,\mu)+ G_{11}^\dagger(t,0,\mu) G_{22}(t,0,\mu)]  + \mathcal{O}(\lambda^2)\, ,\notag
\end{align}
where we made use of the fact that $G_{22}(-\mu)=G_{11}(\mu)$ and the disconnected pieces are of order $\mathcal{O}(\lambda^2)$. The nucleon correlators receive additional connected contributions but, as in the standard case, there are no disconnected pieces. The proton propagator reads
\begin{align}
\langle p(t) p(0) \rangle =& \epsilon_{abc} \epsilon_{ijk} \left \lbrace G_{22}^{ci}(t,0,\mu) \left( G_{11}^{bj}(t,0,\mu) G_{11}^{ak}(t,0,\mu) -G_{11}^{bk}(t,0,\mu) G_{11}^{aj}(t,0,\mu) \right) \right. \notag \\
&- G_{21}^{cj}(t,0,\mu) \left(G_{12}^{bi}(t,0,\mu) G_{11}^{ak}(t,0,\mu) - G_{12}^{ai}(t,0,\mu) G_{11}^{bk}(t,0,\mu)  \right) \\
&+ \left. G_{21}^{ck}(t,0,\mu) \left(G_{12}^{bi}(t,0,\mu) G_{11}^{aj}(t,0,\mu) - G_{12}^{ai}(t,0,\mu) G_{11}^{bj}(t,0,\mu) \right) \right \rbrace \, . \notag 	
\end{align}
We performed simulations and measurements of the spectrum for two parameter sets: $\beta=~4.0,$ $m=~0.05,$ $\lambda=~0.005$ and $\beta=3.5,$ $m=0.01,$ $\lambda=0.001$ both on $16^3\times 24$ lattices. We choose the ration $\frac{\lambda}{m}=0.1$ to keep the correlator contributions from the pion source terms small, as we do not include disconnected pieces in our measurements. We are not able to dismiss the fact that the disconnected pieces could have significant influence on the mesons masses however we checked the significance of the additional connected contributions. Also the additional connected pieces are of order $\mathcal{O}(\lambda^2)$ and their omission in the measurements only lead to neglect able changes in the nucleon and $\pi^0$ correlators. Therefore, we conclude that also the disconnected pieces of the the correlators should only lead to minor corrections.
\subsection{Spectroscopy at vanishing isospin chemical potential}
So far we have used two different parameter sets in our simulations. In both cases simulations were performed on a $16^3\times 24$ lattice. Lattice parameters were
\begin{itemize}
	\item[(1)] $\quad \beta=4.0,$ $m=0.05,$ $ \lambda=0.005$
	\item[(2)] $\quad \beta=3.5,$ $m=0.01,$ $ \lambda=0.001$.
\end{itemize}
We started our analysis by extracting the vacuum masses of the pion, the rho meson and the nucleons using the operators stated in table~\ref{tab-1}.
\begin{table}
\small
\centering
\begin{tabular}{ l c r }
			\hline
  			 & $(1)$ & (2) \\ \hline
  			$am_\pi$ & 0.66(1) & 0.2597(1) \\
  			$am_\rho$ & 1.04(3) & - \\
  			$am_N$ & 1.64(5) & - \\
 			\hline 
		\end{tabular}
		\caption{Vacuum masses of the pion, rho meson and the nucleons for the two parameter sets used.}
		\label{tab-2}
\end{table} \noindent The results for the masses are given in table~\ref{tab-2}. The masses for the rho meson from channel II and III were compatible within errors. Unfortunately,for parameter set 1 the ration of $m_\pi/m_\rho=0.635$ is much larger than the physical value of $m_\pi/m_\rho=0.18$, indicating our pion is very heavy. Therefore, we chose to investigate a second parameter set. Even though we managed to get a lower pion mass in lattice units for parameter set 2 the signal to noise ratio for all other correlation functions were too bad to extract $\rho$ or nucleon masses. Thus, we are not able to make any statements about the behavior of the $\rho$ meson, the nucleons or the physical pion mass.

\subsection{Spectroscopy at finite isospin chemical potential}
\begin{figure}[tp]
\centering
\subfigure[parameter set 1: $ \beta=4.0,$ $m=0.05,$ $\lambda=0.005$]{\includegraphics[width=0.475 \textwidth]{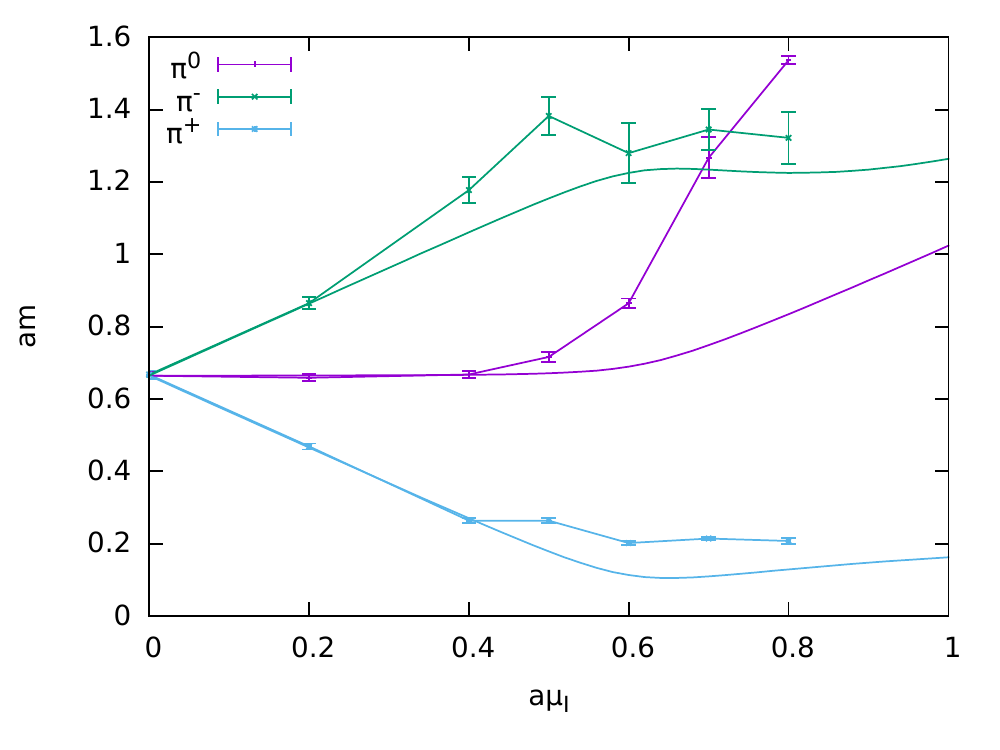}}\hfill
\subfigure[parameter set 2: $\beta=3.5,$ $m=0.01,$ $\lambda=0.001$]{\includegraphics[width=0.475 \textwidth]{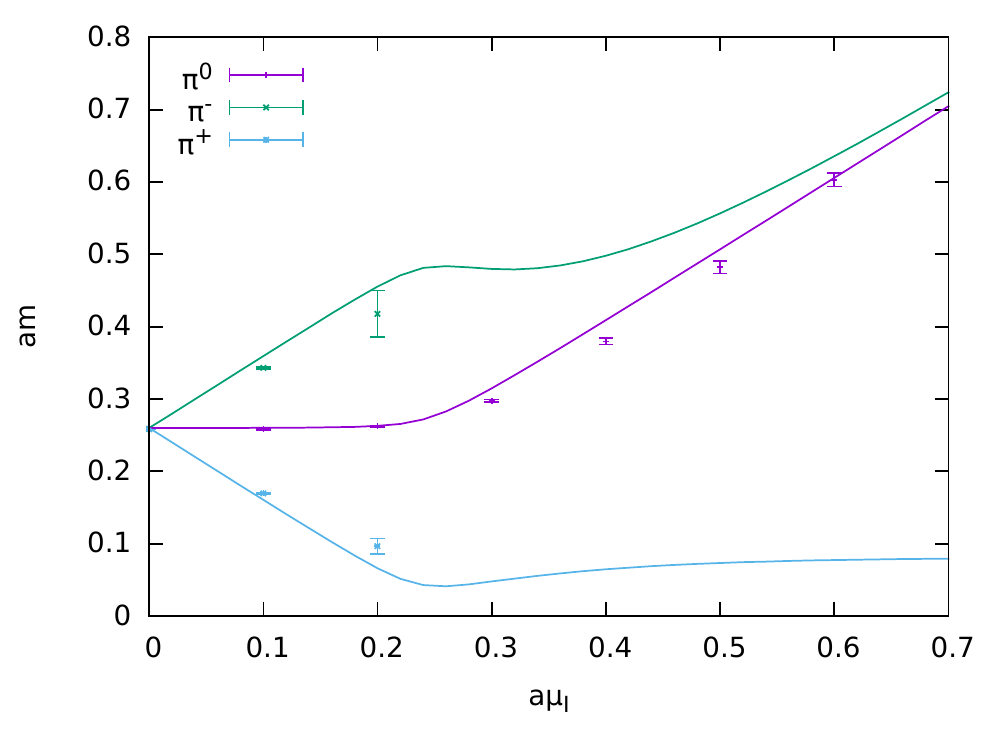}}
\caption{Pion masses in dependence of the isospin chemical potential for both parameter sets. The solid lines are predictions from leading order chiral perturbation theory taking into account the finite value of $\lambda$.}
\label{pic-pions_mu}
\end{figure}

\begin{figure}[tp]
\centering
\subfigure[comparison of $\rho$ and pion masses]{\includegraphics[width=0.475 \textwidth]{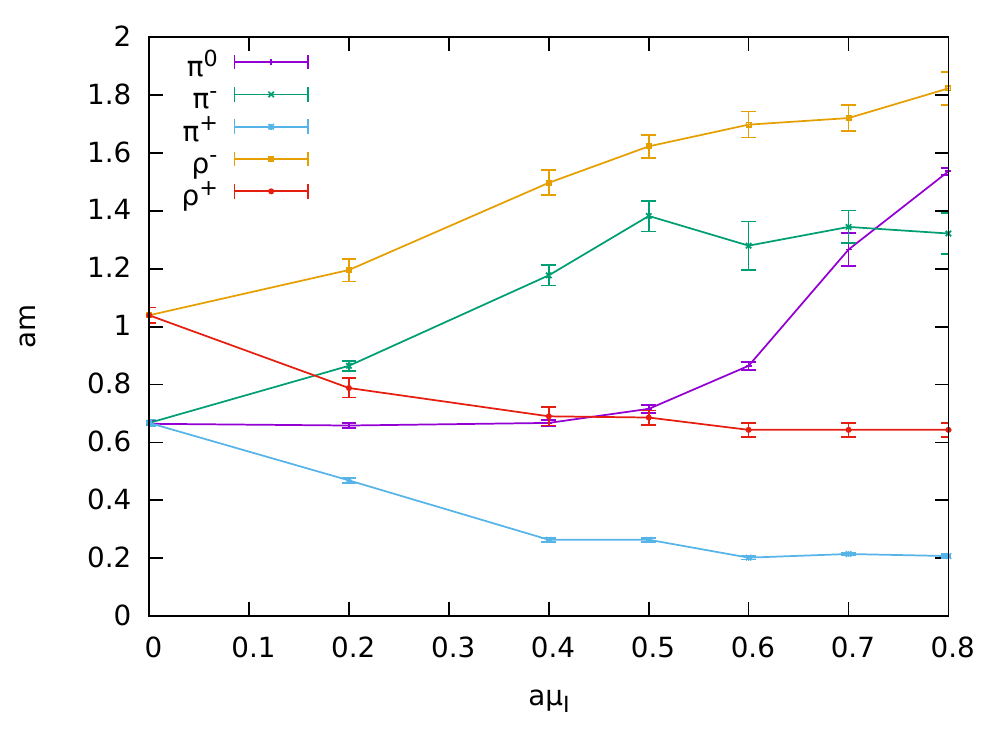}}\hfill
\subfigure[baryon masses]{\includegraphics[width=0.475 \textwidth]{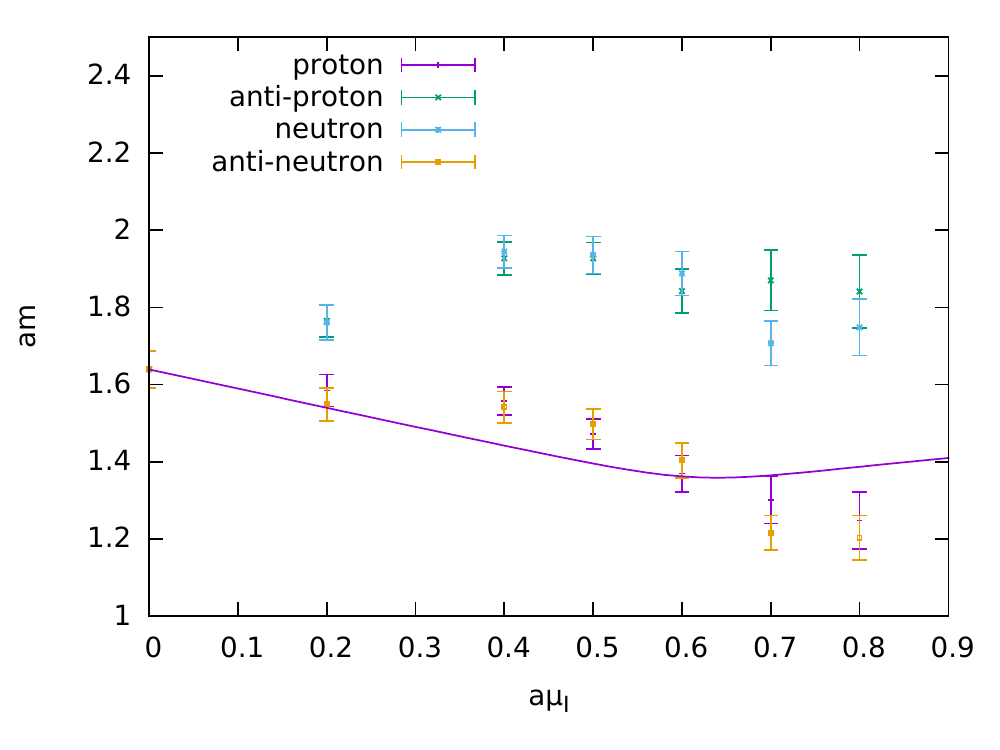}}
\caption{Meson and baryon masses in dependence of the isospin chemical potential for parameter set 1. The solid line is a prediction from leading order baryon chiral perturbation theory.}
\label{pic-mesbar_mu}
\end{figure}

In the vacuum masses of particles and their anti-particles are equal. Therefore, particle correlators on a finite lattice with (anti-)periodic boundary conditions in the time direction always show some symmetry. Depending on the quantum numbers of the particular particle the correlators are either symmetric or anti-symmetric. When working with staggered fermions, it is a little more complicated as staggered lattice operators mix continuum states with different quantum numbers. This may lead to the fact that one finds different symmetries for the correlator parts defined for lattice points with even and odd time coordinates \cite{Altmeyer1993}. By introducing a chemical potential for the quarks we prefer particle over anti-particles or in other words particles propagating forward in time over particles propagating backwards in time. This results in the loss of symmetry in the correlation functions. Introduction of an isospin chemical potential has the same effect as we prefer particles with positive isopin over particles with negative isospin, i.e. up-quark over down-quark but also up-quark over anti-up-quark. For example the correlation function of the $\pi^+$ particle in the vacuum is given by
\begin{equation}
	C(t)=(-1)^t A \left(e^{- m_\pi t} + e^{-m_\pi (N_t-t)} \right) = (-1)^t 2 A \cosh\left(m_\pi \left(t-\frac{N_t}2 \right)\right) \, , \label{corr_sym}
\end{equation}
where the masses of the forward and backward propagating masses are the same: the vacuum pion mass. By introducing the isospin chemical potential we prefer the $\pi^+$ over the $\pi^-$ and the correlation function now reads
\begin{equation}
	C(t, \mu_I) = (-1)^t A \left(e^{- m_{\pi^+}(\mu_I) t} + e^{-m_\pi^-(\mu_I) (N_t-t)} \right) \, , \label{corr_aysm}
\end{equation}
\begin{figure}[tp]
\centering
\subfigure[$a\mu_I=0$]{\includegraphics[width=0.475 \textwidth]{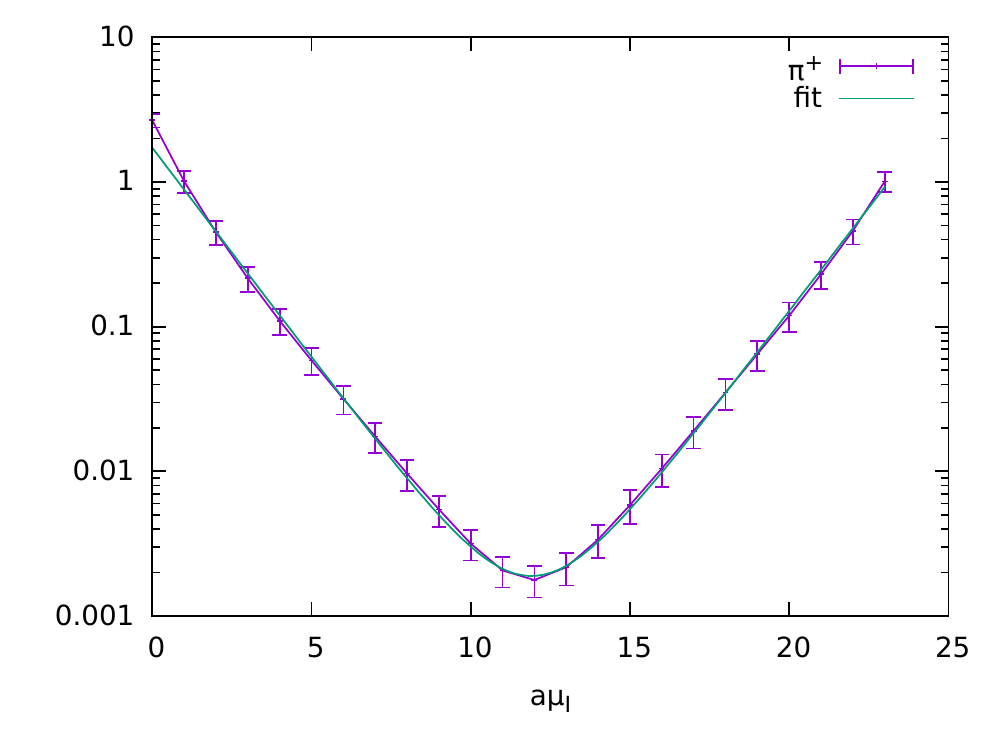}}\hfill
\subfigure[$a\mu_I=0.1$ ]{\includegraphics[width=0.475 \textwidth]{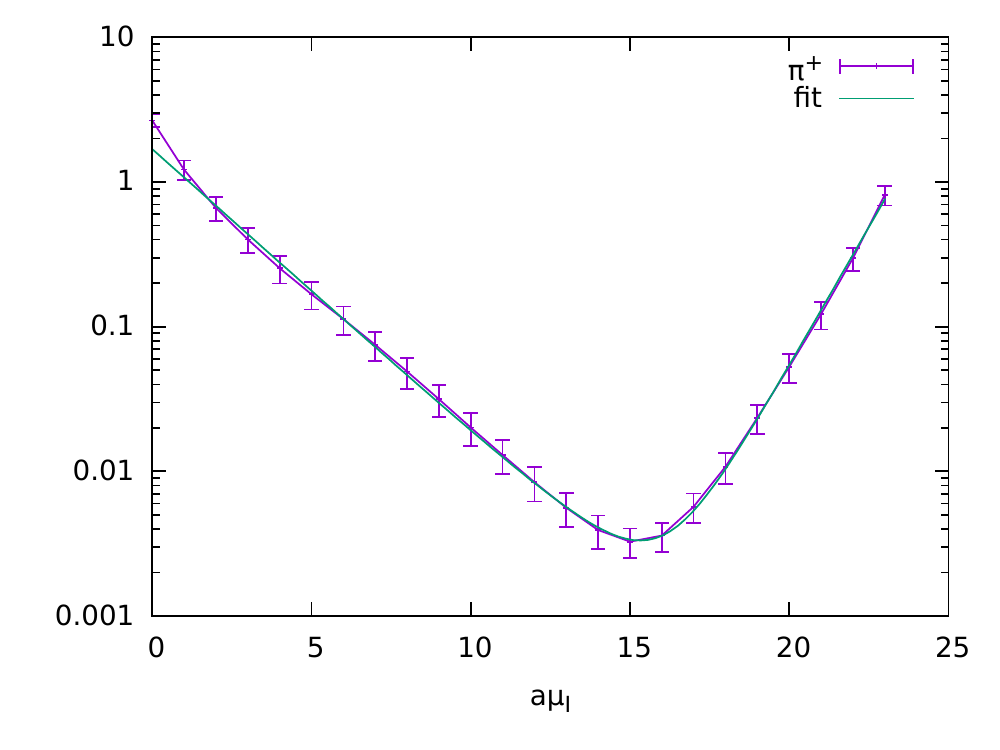}} 
\caption{$\pi^+$ correlation functions for $\beta=4.0,$ $m=0.05$, $\lambda=0.005$. Left: correlator at $a \mu_I=0$ and a fit to eq.~\eqref{corr_sym}. The correlator is symmetric around $N_t/2=12$. Right: correlator at $a \mu_I=0.1$ and a fit to eq.~\eqref{corr_aysm}. The symmetry around $N_t/2$ is lost.}
\label{pic-corrs}
\end{figure}\noindent with a splitting between the masses for $\pi^+$ and $\pi^-$ depending on $\mu_I$. The correlator for the $\pi^0$ will always be symmetric as the $\pi^0$ has isospin zero. However, depending on $\mu_I$ its mass my also differ from the vacuum pion mass. Figure~\ref{pic-corrs} shows the $\pi^+$ correlation function for vanishing isospin chemical potential (a) and for an isospin chemical potential of $a \mu_I=0.1$ (b). The mass splitting of the pions at finite $\mu_I$ is clearly visible in the asymmetry of the correlation function in (b).\\
Predictions for the pion mass splitting can be computed with chiral perturbation theory \cite{Kogut2000477}. The prediction for the dispersion relation of the pions away from the phase with a pion BEC is
\begin{align}
 \pi^+:& \quad E= \sqrt{\vec p^2 +m_\pi^2} - \mu_I \notag \,,\\
 \pi^-:& \quad E=\sqrt{\vec p^2 +m_\pi^2} + \mu_I \, , \\
 \pi^0:& \quad E=\sqrt{\vec p^2 +m_\pi^2} \, . \notag
 \end{align}
 The dispersion relation explicitly show the splitting of the different pion modes according to their isospin charge. When the isospin chemical potential is $\mu_I^c=m_\pi$ we can excite $\pi^+$ particles with $\vec p =0$ essentially for free. This is exactly the critical point where the condensation of $\pi^+$ to a BEC sets in. In our simulations we will never find exactly massless pions as the pion source term acts an infrared cutoff also for the pion mass. We can nevertheless include the pion source term into the chiral perturbation theory calculations to make predictions for the pion spectrum at finite isospin density for finite values of $\lambda$. Again, this is in agreement with the phase diagram of isospin QCD: the introduction of the pion source changes the second order phase transition to the pion BEC phase to an analytic crossover.\\
 Figure~\ref{pic-pions_mu} show the pion masses and mass splittings in dependence of the isospin chemical potential for both parameter sets used in the simulations.
 The solid lines are predictions from leading order chiral perturbation theory, taking into account the finite values of $\lambda$. For the parameter set 1 we find a rather good agreement of our results with the predictions from chiral perturbation theory up to $a \mu_I\approx 0.4$. For higher isospin chemical potential we find that the masses for the $\pi+$ and $\pi^-$ deviate slightly from the predictions. Still, the curves for the charged pion masses level of in the pion condensation phase as predicted. The $\pi^0$ mass starts to deviate from its predicted value also around $a \mu_I\approx 0.4$. Even though chiral perturbation theory predicts an increase of the pions mass, proportional to $\mu_I$, the masses from our data and the slope of the resulting curve are significantly too large. For parameter set 2 the situation is different. We find an excellent agreement of the predicted and measured values of the $\pi^0$ mass for all values of $\mu_I$, even deep inside the pion condensation phase. The agreement for the charged pion masses is also very good, however we only have very few data points for the charged pion masses, as we unfortunately loose signal for the charged pion correlators already for quite small values of the isospin chemical potential. We are not sure, why the $\pi^0$ mass for parameter set 1 starts too deviate significantly from the predictions. Comparison of the situation with parameter set 2 could indicate that is due to the lager pion mass. However, it is still unclear why the deviations are much lager for the $\pi^0$ than for the charged pions.\\
We extracted masses from all correlation functions of operators in table~\ref{tab-1}. As already mentioned above we were not able to extract any masses, except for the pions, for parameter set 2. For parameter set 1 we were able to measure correlation functions for all channels, giving use access to the masses of the $\rho$ mesons and the nucleons. Figure~\ref{pic-mesbar_mu} shows the measured meson and baryon masses in dependence of the isospin chemical potential. The masses of the $\rho$ meson seem to run parallel to the pion masses for all values of $\mu_I$ investigated. This is expected outside the pion condensation phase, as we expect the excitation energy of a particle to be
\begin{equation}
 	E = \sqrt{\vec p ^2 + m^2} +I \mu_I \, ,
 \end{equation} 
where $I$ is the isospin charge of the particle. The almost constant value of the $\rho^+$ inside the phase with non-vanishing $\langle \pi \rangle$ could indicate that the possible formation of a $\rho$ condensate is delayed to higher $\mu_I$. On the other hand, we could have simply reached a lower boundary for $m_\rho(\mu_I)$ that comes with introduction of the pion source term. This has to be investigated in the future by performing a $\lambda \to 0$ extrapolation. In case of the nucleons, we find a near perfect degeneracy of neutron and anti-proton and respectively the anti-neutron and the proton, which is expected as the only difference in our setup between the neutron and the anti-proton is the baryon number of the particles and all our calculations are at vanishing baryon chemical potential. Outside the BEC phase we find a near linear behavior of the nucleon masses. Inside the pion condensation phase, the mass of the neutron and anti-proton seem to be almost constant while the mass of the proton and anti-neutron still seem to decrease further. In contrast a prediction from baryon chiral perturbation theory \cite{PhysRevLett.86.592} is that the proton mass should not decrease further but increase again, as the protons are repelled by the pion condensate. We have not enough data yet to make a definite statement about the fate of the proton inside the pion condensate and clarifying this issue will be an interesting task for further research.

\section{Summary}
We have presented our first results for QCD at finite isopsin density and in particular for the meson and baryon spectrum at finite $\mu_I$. We found that the introduction of a finite pion source term leads to additional terms in the correlation functions for all particles containing not only one single quark species. This makes spectroscopy more challenging, especially in the case of mesons where the pion source term leads to the appearance of disconnected contributions to the correlators of all mesons. So far we only investigated the connected pieces of the correlators. Yet, we do believe that masses extracted from the connected pieces are already a good approximation to the full answer, as the disconnected pieces should only contribute very little in the parameter range we have chosen. Thus, we were able to extract pion, $\rho$ and nucleon masses over a large range of $\mu_I$ until deep inside the pion condensation phase. Unfortunately, the parameter sets we used for our investigations so far are not ideal. The pion in the first parameter set is much heavier as in nature and the second parameter set shows a very bad signal-to-noise ratio for most correlators. Therefore, we will need to find better parameter sets for our future simulations. Further, we will need to extrapolate our results to vanishing pion source to get physically relevant masses.

\bibliography{lattice2017}

\end{document}